\documentclass{aa}

\usepackage{graphicx}
\usepackage{psfig}

\newfont{\vssn}{cmss10 scaled 1050}
\newfont{\vsss}{cmss10 scaled 450}

\newcommand{\as}[2]{$#1''\,\hspace{-1.7mm}.\hspace{.1mm}#2$}
\newcommand{\am}[2]{$#1'\,\hspace{-1.7mm}.\hspace{.0mm}#2$}

\newcommand{\HI}{\mbox{H\,{\sc i}}}
\newcommand{\HIit}{\mbox{H\hspace{0.155 em}{\footnotesize \it I}}}
\newcommand{\HII}{\mbox{H\,{\sc ii}}}
\newcommand{\Jykms}{\mbox{\rm Jy~km\,s$^{-1}$}}
\newcommand{\kms}{\mbox{\rm km\,s$^{-1}$}}

\newcommand{\Msun}{\mbox{${M}_\odot$}}
\newcommand{\MsunLBsun}{\mbox{${M}_{\odot}$/$L_{\odot,B}$}}

\newcommand{\sbb}{mag/$\sq\arcsec$}

\def\H1{\ion{H}{i}}

\def\sbu{${\rm mag\,\,arcsec^{-2 }}$\ }
\def\p25{P$_{25}$}
\def\e25{E$_{25}$}
\def\kato{\rule[-1.25ex]{0cm}{1.25ex}}
\def\pano{\rule[0.0ex]{0cm}{2.5ex}}

\def\cg{mag kpc$^{-1}$}
\newcounter{qub}
\begin{document}

\title{A study of H{\bf \Large I}-selected galaxies in the Hercules cluster}

\author{J. Iglesias-P\'{a}ramo
      \inst{1}
      \and
W. van Driel
	\inst{2}
	\and
P.-A. Duc
      \inst{3}
	\and
P. Papaderos
	\inst{4}
	\and
J.M. V\'{\i}lchez
	\inst{5}
	\and
V. Cayatte
	\inst{2}
	\and
C. Balkowski
	\inst{2}
	\and
K. O'Neil
            \inst{6}
	\and
J. Dickey
	\inst{7}
	\and
H. Hern\'andez
             \inst{6}
	\and
T.X. Thuan
	\inst{8}
          }
\offprints{jorge.iglesias@oamp.fr}
\institute{Laboratoire d'Astrophysique de Marseille, Traverse du Siphon - 
Les Trois Lucs, 13376 Marseille, France\\
              \email{jorge.iglesias@oamp.fr}
         \and
Observatoire de Paris, GEPI, CNRS-UMR 8111 and Universit\'{e} Paris 7, 92195
Meudon Cedex, France\\
\email{wim.vandriel@obspm.fr,veronique.cayatte@obspm.fr, chantal.balkowski@obspm.fr}
	\and
CNRS URA 2052 and CEA, DSM, DAPNIA, Service d'Astrophysique, 91191
Gif-sur-Yvette Cedex, France \\
	\email{paduc@cea.fr}
         \and
Universit\"{a}ts-Sternwarte, Geismarlandstrasse 11, 37083 G\"{o}ttingen, Germany\\
\email{papade@uni-sw.gwdg.de}
	\and
Instituto de Astrof\'{\i}sica de Andaluc\'{\i}a (CSIC), Granada, Spain\\
\email{jvm@iaa.es}
               \and
Arecibo Observatory, HC3 Box 53995, Arecibo, Puerto Rico 00612, U.S.A. \\
             \email{koneil@naic.edu, hhernand@naic.edu}
	\and
Department of Astronomy, University of Minnesota, 116 Church Street SE,
Minneapolis, MN 55455, U.S.A.\\
\email{john@astro.umn.edu}
	\and
Astronomy Department, University of Virginia, Charlottesville, VA 22903, U.S.A.\\
\email{txt@astro.virginia.edu}
            }

        \date{\it Submitted to Astronomy and Astrophysics -- second revised version  17.2.2003}


\abstract{\rm
{The present study is aimed at a sample of 22 galaxies detected in the blind VLA \HI\ survey
of the Hercules cluster by Dickey (1997), 18 of which were selected on an \HI\ line width smaller 
than 270 \kms\ and 4 others with only tentative optical counterparts on the Palomar Sky Survey.
Sensitive single-dish \HI\ line spectra were obtained for 20 of them, and for one (47-154) 
the VLA detection was not confirmed.
Optical surface photometry was obtained of 10 objects, for 8 of which optical spectroscopy 
was obtained as well. Based on various selection 
criteria, two (ce-143 and ne-204) can be classified as dwarfs. The objects of
which optical observations were made show star formation properties similar to those of published 
samples of actively star forming galaxies, and approximately half of them
have properties intermediate between those of dwarf galaxies and low-luminosity disc galaxies. 
No optical redshifts could be obtained for two of the galaxies (sw-103 and sw-194) and their 
physical association with the \HI\ clouds detected at their positions therefore remains uncertain. 
Unique among the objects is the 
Tidal Dwarf Galaxy ce-061 in a tail of the IC 1182 merger system.
} %
}

\maketitle

\markboth {Iglesias-P\'{a}ramo et al.}{A study of HI-selected galaxies in the Hercules cluster}


\section{Introduction \label{S1} }  
It is well known from numerous studies based on observations and simulations that the 
environment plays a fundamental role in the evolution of bright galaxies, via galaxy-galaxy 
interactions and/or interactions of galaxies with the intergalactic medium. 
The present work is part of an ongoing study of the properties of \HI-selected galaxies in 
clusters, for which the results on dwarfs in the Hydra cluster have already been published (Duc et 
al. 1999, 2001a; hereafter Paper I and II, respectively). The most remarkable result is the existence 
of dwarfs with an oxygen abundance significantly higher than expected from the 
luminosity-metallicity relation for field dwarf galaxies.

In this paper we present single-dish 21 cm \HI-line as well as optical imaging and spectroscopy 
observations of a sample of galaxies in the Hercules cluster, selected from the objects detected 
in the blind VLA \HI\ line survey of the cluster by Dickey (1997).
One of our aims is to examine which of them can be considered dwarf galaxies. 

The Hercules supercluster is one of the most massive structures in the nearby Universe (Freudling 
1995). It appears to be a collection of three clusters, gravitationally bound, but far from dynamical 
relaxation: Abell~2151, classified as richness class 2, and Abell~2147 and Abell~2152, both 
classified as richness class 1 (Barmby \& Huchra 1998). In a previous single-dish study of the \HI\ 
properties of a sample of galaxies in the Hercules supercluster, Giovanelli et al. (1981) found a 
strong deficiency in the \HI\ mass-to-optical luminosity ratio of galaxies in the Abell~2147 cluster, 
while an almost normal, or mildly deficient, ratio was found for the galaxies in Abell~2151, the 
richest and densest of the three clusters. We have assumed a distance of 150 Mpc to the Hercules 
supercluster, based on an average redshift of 11,050 \kms\ for the cluster spirals (see D97) and a 
Hubble constant of 75 km s$^{-1}$ Mpc$^{-1}$. 

The paper is organized as follows: the sample selection and \HI\ and optical observations are 
described in section~2. The optical and \HI\ properties of the sample galaxies are described 
insection~3, and a discussion of the results is presented in section~4. Comments on individual objects are 
given in an Appendix.

\section{Observations}  
\subsection{Sample selection} 
Our study of the Hercules cluster is aimed at \HI-selected galaxies, as well as at reported
\HI\ clouds without optical 
counterparts on Digital Sky Survey (DSS) images, selected from the blind VLA 21~cm \HI\ line 
survey of the cluster by Dickey (1997, hereafter D97) and reobserved by us at Arecibo in order to 
confirm the detections and to obtain \HI\ line profiles with a better velocity resolution. 
Our failure to confirm the \HI\ clouds without optical counterparts indicate that these 
were spurious, as described in van Driel et al. (2003). 

For our studies of \HI-selected dwarf galaxies in the Hydra cluster (Papers I and II) we 
used the VLA \HI\ study by McMahon (1993), which has a velocity resolution of 42 \kms, similar
to that of the D97 survey. We selected on \HI\ line widths at 20\% of 
peak flux density value, $W_{20}$, smaller than 130~\kms, as objects with such narrow lines are 
good dwarf candidates. Of the 20 selected objects only 4 were found not to be dwarfs.

We could not apply such an effective \HI\ line width criterion aimed at selecting dwarfs to the D97 data, 
however. As the Hercules cluster is about three times more distant (150 Mpc) than the Hydra cluster 
(45 Mpc), the \HI\ profiles are correspondingly weaker and their widths more uncertain.
The D97 \HI\ mass detection limit of about 5 10$^8$ \Msun\ allows the detection of only the most gas-rich 
dwarf systems, 
while we estimate that the uncertainty in the FWHM line widths, $W_{50}$, is about 100 \kms\ for 
the fainter \HI\ detections, following Fouqu\'e et al. (1990). We raised the cut-off value for
$W_{50}$ to 270~km~s$^{-1}$ for the Hercules cluster, thereby excluding only the most massive, 
inclined spiral systems.

All four fields of the D97 VLA study (three of which -- ne, ce and sw -- are located in Abell~2151, 
while the forth -- 47 -- is centred on Abell~2147) were covered in our Arecibo single-dish \HI\ study. 
Of the total of 25 galaxies with optical counterparts on the Digital Sky Survey (DSS) and showing line widths 
smaller than 270~\kms\ 
detected in the 4 fields by D97, we observed the following 18 in the \HI\ line at Arecibo: ne-112, 
ne-142, ne-178, ne-204, ne-208, ne-240, ce-042, ce-048, ce-060, ce-061, ce-143, ce-176, ce-200, 
sw-103, sw-222, 47-138, 47-166 and 47-211. The 7 others were ruled out because they are very 
likely face-on spirals: ne-169, ne-222, ne-264, ce-122, ce-166, sw-159 and 47-030. In addition, we 
included in our list four galaxies showing line widths larger than 270~\kms\ but with only tentative 
optical counterparts on the DSS (ne-250, ne-398, sw-194 and 47-154) in order to confirm the \HI\ 
detections and to verify whether the optical associations are real or not. 
A summary of the observations obtained for our survey, together with the centre positions and $W_{50}$
line widths from D97, is given in Table 1.

For our optical imaging and spectroscopic observations, only the central (ce) and southern (sw) 
fields could be covered.

\subsection{\HIit\ line observations} 
We made our \HI\ line observations of the 22 \HI-selected galaxies in the Hercules cluster with the 
refurbished 305 m Arecibo Gregorian radio telescope in May and June 2002. For further technical 
details on the observations and the data reduction we refer to van Driel et al. (2003). 

The total net integration time (on+off) was on average 70 minutes per source, depending on the line 
strength, from 40 minutes for the strongest lines to 110 minutes for the weakest signals, in ne-398 
and 47-154. The velocity coverage is about 2500 \kms, the velocity resolution is 1.3 \kms, and the 
telescope's HPBW at 21 cm is \am{3}{4}$\times$\am{3}{6}. For the telescope's pointing positions 
the centre coordinates of the VLA \HI\ sources as given in D97 were used (see Table~1 of D97). 
The data were reduced using IDL routines developed at Arecibo Observatory. A first-order baseline 
was then fitted to the data and the velocities were corrected to the heliocentric system, using the 
optical convention. All data were boxcar smoothed to a velocity resolution of 9.1 \kms\ for further 
analysis, while the data of ne-398 and 47-154 were smoothed to 19.5 \kms. 

\subsection{Optical observations} 
Our optical observations are limited to objects in the ce and sw fields of the D97 survey. 
Of all 10 sample galaxies in these two fields we obtained deep CCD images, and 
low-to-medium resolution spectra for the 8 among these with the brightest optical counterparts
(see Table 1). 

\subsubsection{Optical imaging} 
$B$, $V$ and $i$-band images were taken for most objects in our optical sample, except ce-143 
and sw-222, for which only $V$ and $i$ imaging could be obtained. Observations were carried out 
with the Wide Field Camera attached to the prime focus of the 2.5m Isaac Newton Telescope (INT) of the 
Observatorio del Roque de los Muchachos, in Spain, on June 5, 1999 and April 26, 2000. Both 
fields were observed under photometric conditions. The WFC consists of a science array of four 
thinned AR coated EEV 4k$\times$2k devices, plus a fifth used for autoguiding. The pixel scale is 
0.33 arcsec~pixel$^{-1}$, which gives a total field of view of about $34'\times$34$'$. Given the 
particular arrangement of the detectors, an area of about $11'\times$11$'$ is not usable at the top 
right corner of the field. 

The global accuracy of the photometry is about 0.10~mag. For the $i$-band frames, the accuracy is 
slightly poorer due to residual fringing and the accuracy of the colour term $\delta (V-i)$ is 
about 0.15~mag. Although a Sloan-Gunn $i$ filter was used instead of Cousin $I$, the 
reported $I$ magnitudes correspond to the Cousin system, as photometric standards from  
Landolt (1992) were observed and a linear relationship with a slope unity was found 
between the expected number of counts for each of the filters. The astrometry of the images was 
carried out using USNO guide stars. Detailed $V$-band images as well as $(V-I)$ colour maps 
are shown in Section~3.2. 

\subsubsection{Optical spectroscopy} 
Medium and low-resolution spectroscopy was carried out at the 4.2m William Herschel Telescope 
(WHT) and the 2.5m Nordic Optical Telescope (NOT) at the Observatorio del Roque de los 
Muchachos, in Spain. Table~\ref{logspec} shows the diary of the spectroscopic observations. 
Observations at the WHT were performed during several nights using the double arm spectrograph 
ISIS, with the dichroic splitting the beam set at 5700\AA. For most of the galaxies observed at the 
WHT, the CCD set-up was two 1k$\times$1k Tektronix per arm. The gratings used were R300B 
and R316R, giving nominal dispersions of 1.54\AA~pix$^{-1}$ and 1.49\AA~pix$^{-1}$ for the 
blue and red arms, respectively. The corresponding spectral coverages were 3735--5311\AA\ and 
6118--7643\AA, respectively, and the spatial scale in both detectors was 
0.36~arcsec~pix$^{-1}$. Note that ce-143 was observed with a different set-up, using a EEV 
2k$\times$4k detector on the blue arm, giving a nominal dispersion of 0.86\AA~pix$^{-1}$ and a 
spatial scale of 0.2~arcsec~pix$^{-1}$. In all cases, the slit width was set to match the seeing, 
about \as{1}{0} for most of the objects. Observations at the NOT were taken using the faint object 
spectrograph ALFOSC, with a 2k$\times$2k LORAL detector and Grism \#4, giving a 
nominal dispersion of 3.3\AA~pix$^{-1}$. The total wavelength coverage with this set-up 
was 3200--9100\AA, and the spatial scale was 0.18~arcsec~pix$^{-1}$. 

The slit was always centred on the galaxy nucleus and positioned along the position angle of the
major axis, as listed in Table~\ref{logspec}. Although spectro-photometric standard stars were 
observed for flux calibration, only the relative fluxes of the emission lines are reliable, 
since several nights were non-photometric. 

The emission lines were measured with the SPLOT package running on IRAF. For each emission 
line five independent measures were performed, and the adopted fluxes and errors are, respectively, the 
average and standard deviation of the five measures. For ce-042, ce-200 and sw-222 the intensities 
of the [N{\sc ii}] doublet and the H$\alpha$ line were determined by fitting the non-resolved triplet 
with three gaussians. Similarly, the fluxes of the [S{\sc ii}] lines of sw-222 were obtained by fitting 
the doublet with two gaussians. Larger errors resulted from this deblending process. No correction 
was made for line absorption in the Balmer lines. Only ce-200 shows an absorption feature at 
H$\beta$, for which we simply measured the intensity from the base of the emission line. For some 
of the spectra obtained at the WHT, we had to rescale the red part in order to get the same 
continuum levels at both sides of the gap, since it is well established that the continuum of H{\sc ii} 
regions/galaxies is smooth at these wavelengths. All lines were dereddened using the extinction 
coefficient derived from the Balmer decrement H$\alpha$/H$\beta$ and assuming a theoretical 
value for the intrinsic line ratio of 2.89 (Brocklehurst 1971). No extinction correction was applied 
to galaxies for which the observed H$\alpha$/H$\beta$ flux ratio was consistent with this value, 
within the errors.

\section{Results}  
\subsection{\HIit\ properties} 
The Arecibo \HI\ spectra of the 20 galaxies for which we could obtain sensitive spectra are shown 
in Figure~\ref{arecibospect}, smoothed to a resolution of 9.1 
\kms\ for most objects, and to 19.5 \kms\ for the weak line signal of ne-398 and for the undetected 
object 47-154. Not shown are the spectra of sw-194 and 47-138, for which no sensitive \HI\ 
observations could be obtained due to the proximity of strong continuum sources. 

Besides the VLA data of D97, on which the present study is based, published \HI\
detections (see Appendix A) were only found for the merger system IC 1182, which contains 
the ce-061 tidal dwarf galaxy in one of its tails.

We compared (Table~\ref{hiressum}) the Arecibo global \HI\ profile parameters to those of the D97 
VLA observations, which have a synthesized beam size (HPBW) varying from 20$''$$\times$21$''$ 
to 26$''$$\times$29$''$, a velocity resolution of 44.2 \kms\ (degraded to 88 \kms\ for the 
determination of the profile parameters), an rms noise of about 0.13 mJy/beam per channel 
map at the centre of the primary beam, and a pixel size of 6$''$$\times$6$''$, i.e. about 24 
pixels per synthesized beam. 
Values in brackets indicate the 4 Arecibo profiles estimated to be significantly confused 
by line emission from nearby galaxies: ne-240, ce-048, ce-060 and ce-200 (see Appendix~A for 
comments on these objects). Listed in the columns are the following data; note that all radial 
velocities in this paper, 
both optical and \HI, are in the heliocentric system, using the conventional optical definition 
(V=c($\lambda$-$\lambda_0$)/$\lambda_0$):  (1)  the galaxy's name, from D97, (2) the centre 
velocity of the VLA profile, $V_{HI}$, (3) the width of the VLA profile at 50\% of the maximum 
flux density, 
$W_{50}$, (4) the $I_{H}$ integrated VLA \HI\ line flux (see the description below), (5) the 
$I_{ext}$ integrated VLA \HI\ line flux (see the description below), (6) the centre velocity of the 
Arecibo \HI\ profile, $V_{HI}$, (7) the width of the Arecibo profile at 50\% of the maximum flux 
density, $W_{50}$, (8) the width of the Arecibo profile at 20\% of the maximum flux density, 
$W_{20}$, (9) the integrated Arecibo \HI\ line flux, $I_{HI}$ and (10) the rms noise levels of the 
Arecibo spectra. 

As it is in principle not straightforward to determine the integrated \HI\ line profiles parameters
of faint objects from interferometric data, four different methods were used in D97. 
We converted the \HI\ masses listed in D97 to integrated \HI\ line 
fluxes assuming the cluster distance of 110.5 Mpc adopted in D97. The characteristics of these 
methods are as follows: $I_{H}$ is the line flux measured by integrating the spectra over the group 
of contiguous pixels above threshold, $I_{ext}$ is the line flux integrated over a larger area, 
estimated by statistical tests to contain the total line emission, while $I_{peak}$ and $I_{int}$ are 
obtained by fitting a two-dimensional gaussian to the velocity-integrated \HI\ column density map, 
where $I_{peak}$ corresponds to the line flux within the central beam area and $I_{int}$ to the 
integrated flux of the gaussian. The latter is notoriously unstable. 

Although observations with a single dish telescope like Arecibo result in only one spectrum per 
pointing position and the derivation of integrated \HI\ profile parameters is straightforward, these 
profiles depend on the instrument's beam pattern, which can lead to confusion with other objects 
in the beam (see Appendix~A), and single-dish data are more sensitive to RFI than interferometric data. 

We estimated the uncertainties, $\sigma_{V_{HI}}$, in the central \HI\ velocities of the Arecibo 
line profiles, following Fouqu\'e et al. (1990): 
\begin{equation} 
\sigma_{V_{HI}} = 4 R^{0.5}P_{W}^{0.5}X^{-1}\,\, [km s^{-1}] 
\end{equation} 
where $R$ is the velocity resolution in \kms, $P_{W}$=($W_{20}$--$W_{50}$)/2 in \kms\ and 
$X$ is the signal-to-noise ratio of a spectrum, which we defined as the ratio of the peak flux 
density and the rms noise. The estimated uncertainties vary from 2 to 14 \kms\ and are on average 
4.6$\pm$2.5 \kms. According to Fouqu\'e et al., the uncertainty in the line widths is 
2$\sigma_{V_{HI}}$ for $W_{50}$ and 3$\sigma_{V_{HI}}$ for $W_{20}$. 

For the 13 objects for which an unambiguous comparison between the integrated Arecibo and VLA 
line fluxes can be made (see Table 1) the average ratio, and its $\sigma_N$ deviation, between the two VLA flux 
determinations made without fitting a gaussian to the source, $I_{H}$ and $I_{ext}$, and the 
Arecibo flux, $I_{HI}$, is $I_{H}$= 0.66$\pm$0.27 $I_{HI}$ and $I_{ext}$= 1.03$\pm$0.47 $I_{HI}$.

We also estimated the uncertainties, $\sigma_{I_{HI}}$, in the integrated line fluxes of the 
Arecibo line profiles following Fouqu\'e et al. (1990), assuming that this formula, which was 
developed for Nan\c{c}ay data, can be applied to Arecibo data as well: 
\begin{equation} 
\sigma_{I_{HI}} = 5 R^{0.5}I_{HI}^{0.5}h^{0.5}X^{-1}\,\, [Jy km s^{-1}] 
\end{equation} 
where $R$ and $X$ are as in equation (1) and $h$ is the peak flux density of the profile, in Jy. For the 
13 abovementioned galaxies the estimated uncertainties 
vary between 5 and 30\% of $I_{HI}$ and are on average ($0.16 \pm 0.08) \times I_{HI}$. 

A comparison of the $W_{50}$ line widths measured with the VLA, which were used for the 
galaxy selection, and at Arecibo (Figure~\ref{w50comp}) clearly shows the effect of the difference 
in velocity resolution, 88 and 9 \kms, respectively. It appears that, on average, the VLA line widths 
have been overestimated by about half a VLA channel width, i.e. 20 \kms. 

Table~\ref{propsum} shows a summary of the \HI\ properties of the sample galaxies as well as 
the absolute $B$ magnitudes of their optical counterparts. Listed in the columns of this Table are the 
following data: (1) the galaxy name, from D97, (2) the central radial velocity 
of the \HI\ profile, $V_{HI}$, (3) the global form of the Arecibo spectra that are not confused, 
where DH denotes a double-horned profile, FT a flat topped one, G a gaussian one and LS a 
lopsided profile, (4) the total \HI\ mass, $M_{HI}$, (5) the relative \HI\ gas content, 
$M_{HI}/L_B$, (6) the FWHM of the \HI\ line, $W_{50}$, and (7) the absolute magnitude in the 
$B$ band, $M_{B}$, from NED. For the \HI-related properties we preferentially used our Arecibo 
spectra, except for cases of confusion with nearby galaxies or a nearby strong continuum source, 
where we adopted the D97 VLA data, using the $I_{ext}$ estimate for the integrated line flux. 

As mentioned above, the reported VLA \HI\ detections of two of the galaxies -- ne-398 and 47-154 
-- were not confirmed at Arecibo. 

\subsection{Optical properties} 
In this section we study the results obtained for the properties of the subsample for which we obtained 
optical spectra and/or imaging. Figures~\ref{field}a and~\ref{field}b show our $V$-band images of 
the ce and sw fields, respectively, with superimposed contours showing the \HI\ clouds 
detected in D97. Figure~\ref{spectra} displays the optical spectra of all galaxies with emission lines and 
Table~\ref{propspec} lists their spectro-photometric data. Uncertainties derived from the 
deblending process for the [N{\sc ii}] and [S{\sc ii}] lines of the galaxies observed at the NOT are 
not included in the Table. 

\subsubsection{Structural properties \label{SurfPhot}} 
In Table~\ref{magcol} we list the absolute magnitudes and colours obtained for the 10 objects of the 
optical subsample, including the two tentative detections sw-103 and sw-194. For the seven 
galaxies which show optical emission lines, the magnitudes are extinction corrected 
using the $C$(H$\beta$) values listed in Table~\ref{propspec}. The magnitudes listed 
in this Table will be used hereafter for this subsample of galaxies. 

For the surface photometry analysis we employed improved versions of the techniques described in 
Papaderos et al. (\cite{P96}a). Before determining the profiles we removed foreground stars and 
background galaxies intersecting the \H1\ galaxies and smoothed all images of a given object to the 
same resolution. Surface brightness profiles (SBPs) were corrected for extinction inside the 
galaxies using the C(H$\beta$) values listed in Table~\ref{propspec}. 

In Table~\ref{Tab1} we summarize the derived photometric properties. Listed in the 12 columns of 
this Table are the following data: (1) the parameters b and q, describing, respectively, the intensity 
distribution of the latter fitting formula near the centre, (3 and 4) the central surface brightness, 
$\mu_{\rm E,0}$, and exponential scale-length $\alpha$, respectively, of the Low Surface 
Brightness (LSB) component, as obtained from linear fits to the exponential regime of each profile 
and weighted by the photometric uncertainty of each point -- note that Eq. 22 in Papaderos et al. 
(1996a) predicts for the LSB component an actual central surface brightness $2.5 \log(1-q)$ mag 
fainter than the extrapolated value $\mu_{\rm E,0}$, listed in Col. 3, (5) the corresponding total 
magnitude of the LSB component as obtained for a pure or a modified exponential distribution, (6-
9) list quantities obtained from profile decompositions; in (8) and (6) are listed, respectively, the 
isophotal radii \e25\ and \p25\ of the LSB component and of the luminosity component in excess of 
it, both determined at the 25\ \sbu\ level, while in (9) and (7) are listed, respectively, the 
corresponding apparent magnitude of these components, $m_{E25}$ and $m_{P25}$,(10) the total 
apparent magnitude, as inferred from the SBP integration out to the last point, $m_{SBP}$, (11) 
the effective radius, $r_{\rm eff}$, and the radius $r_{80}$, which encircles 80\% of the galaxy's 
total flux, and (12) the S\'ersic index $\eta$ resulting from fitting Eq.~5 in Papaderos et al. (1996a) 
to the BSP. 

$V$-band images, $(V-I)$ colour maps, surface brightness and colour profiles of the galaxies are 
presented in Figures~\ref{ce042} to~\ref{sw222}. A common property of all selected galaxies is 
the absence of notable colour gradients. In all cases these do not exceed 0.1 \cg, as also reported for 
many dIs (e.g. Patterson \& Thuan 1996, van Zee 2001). The situation is strikingly different in the 
inner regions ($R^*\la$\e25) of BCDs, where colour gradients of up to $\gamma_+\sim$1.8 \cg\ 
have been observed, like H1034-2558 in the Hydra cluster (Paper I) and other examples in 
Papaderos et al. (\cite{P96}a), Doublier et al. (\cite{Doubl99}) and Cair\'os et al. (\cite{LM01a}). 

For all galaxies except ce-061, an exponential fitting law provides a reasonable approximation to 
the SBPs in their outer low surface brightness regime. For half of the sample galaxies, 
however, inwards extrapolation of this outer exponential slope predicts a higher intensity than 
actually observed. This type of convex profile with an exponential distribution in the outer parts 
and levelling off in the inner part (within 1--3 disc scale lengths) is not rare among low-luminosity 
dwarf ellipticals (e.g. Binggeli \& Cameron \cite{binggeli91}), dwarf irregulars (Patterson \& 
Thuan \cite{PT96}, van Zee \cite{vZ00}), blue LSB galaxies (R\"onnback \& 
Bergvall 1994), near-infrared selected LSBs (Monnier Ragaigne et al. 2003), and has been reported 
in a few blue compact dwarfs (e.g., Fricke et al. 
\cite{Fricke00}; Guseva et al. \cite{Nat01}; Vennik et al. \cite{vennik00}). Note that similar SBPs 
have also been derived for 4  of the \HI-selected dwarfs in the Hydra cluster (H1031-2818, H1031-2632, 
H1032-2722 andH1033-2642; see Paper I). For those systems, following the procedure outlined in 
Guseva et al. (2001), we modelled the LSB component using Eq. 22 in Papaderos et al. (1996a). 

In Figure~\ref{dwarfsdiag} we show $M_{B}$ as function of $\mu_{0,B}$ for the Hercules and Hydra cluster 
galaxies. We used the average $(B-V)$ colour of 0.3 of the other galaxies in the subsample for 
ce-143 and sw-222, and included them in the plot. Also shown are the loci occupied by spiral discs, 
ellipticals and spiral bulges, dwarf Irregulars and dwarf ellipticals from Binggeli (1994), as well as 
the loci occupied by the LSB galaxies from van der Hulst (1998). All Hydra cluster galaxies  and
5 of the 8 Hercules objects lie in the zone occupied by the faintest disc-like galaxies and by dwarf 
systems.

\subsubsection{Star formation activity} 
Not surprisingly for an \HI-selected sample, seven objects out of the eight for which we took 
optical spectra show emission lines, with line ratios typical of H{\sc ii} regions, indicating they are 
star-forming objects. The only quiescent galaxy is ce-060, which shows no sign of ongoing star formation. 
Unfortunately, non-photometric 
weather conditions did not allow us to derive a reliable star-formation rate from the H$\alpha$ 
luminosity. 

In Figure~\ref{salzer} we show the absolute blue magnitudes $M_{B}$  as function of the 
H$\beta$ equivalent widths for the Hercules and Hydra cluster galaxies. For comparison we have 
added the emission line galaxies from Salzer et al. (1989) as well as the upper envelope from 
V\'{\i}lchez (1995). As can be seen from the figure, the Hercules cluster galaxies have 
H$\beta$ equivalent widths normal for their $B$-band luminosities, compared to the Salzer et al. 
sample, whereas the Hydra cluster dwarfs seem to delineate the lower envelope of the locus 
occupied by the Salzer et al. sample.

Broad-band colours can also be used as a diagnostic of star formation, and we show the 
$(B-V)~vs.~(V-I)$ diagram for our sample galaxies in Figure~\ref{colour}. For comparison, 
we have added the ellipticals from Goudfrooij et al. (1994), the spirals from Heraudeau \& Simien 
(1996), the amorphous galaxies from Gallagher \& Hunter (1987) and the nearby dwarfs from 
Makarova (1999) - the latter objects were defined by Sandage \& Brucato (1979) as blue galaxies 
showing centrally concentrated light distributions, with a high \HI\ gas content, low rotation 
velocities and heavy-element abundances typical of Magellanic Irregulars. We notice that all 
7 of our Hercules cluster galaxies with both colours available are concentrated towards the 
bottom left region of the diagram, i.e. the blue part of the sequence delineated by the 
amorphous and dwarf samples, where no ellipticals and only a marginal number of spirals are found. 

\subsubsection{Metallicity}   
The \HI-based selection process resulted in star--forming galaxies hosting \HII\ regions of which 
the metallicity of the ionised gas could be estimated from their emission lines. As the [O{\sc 
iii}]$\lambda$4363\AA\ line was not detected for any of the galaxies, the oxygen abundances were 
computed from the [O{\sc ii}]$\lambda$3727\AA\ and [O{\sc iii}]$\lambda$4959,5007\AA\ 
emission line fluxes, using several abundance calibrations: the $R_{23}$ method calibrated
theoretically by McGaugh et al. (1991), in its parameterised form from Kobulnicky et al. (1999), 
the more recent p-method (Pilyugin 2000,2001) and the independent method of  van Zee et al. (1998),
based on the [N{\sc ii}]/H$\alpha$ line ratio. The first two methods each provide two possible values 
of the metallicity for a given observable. To resolve the degeneracy and choose between the lower and 
upper values, we relied on the strength of the [N{\sc ii}]/[O{\sc ii}] ratio (see van Zee et al. 1998). 
The values of $12 + \log$(O/H) obtained with these three methods, as well as the final adopted values, 
are listed in Table~\ref{oxabun}. 

The criteria used to choose the final adopted abundances listed in column~7 were the following: the 
abundances were preferably derived using the p-method, for consistency with previous work in this 
program (Paper~II). Although the method based on [N{\sc ii}]/H$\alpha$ is not very reliable, 
because of blending problems of the H$\alpha$ and [N{\sc ii}] lines in the low resolution NOT 
spectra, it was preferred for those objects (ce-042 and ce-200, marked with a colon in 
Table~\ref{oxabun}) whose observed line fluxes are not within the validity range of the p-method, 
and which yield inconsistent metallicities: i.e. the oxygen abundance read on the lower branch is 
larger than that read on the upper branch. 

The median oxygen abundance of these 7 Hercules cluster sample galaxies is about one third solar 
($12 + \log$(O/H) = 8.27$\pm$0.31), with a quite large scatter. Figure~\ref{lumi_meta} shows their
metallicities as function of absolute $B$ magnitude; 
the typical uncertainty in the oxygen abundances is indicated by the error bar. For the two galaxies 
without an available $B$ magnitude, ce-143 and sw-222, a $(B-V)$ colour of 0.30 -- the 
median value measured for the other Hercules cluster galaxies -- was assumed. 
For comparison, we added the isolated dwarf Irregulars of Richer \& McCall (1995), the LSBs from 
van der Hulst et al. (1998), the \HI-selected dwarfs from the  Hydra cluster (paper II) and
the tidal dwarf galaxies (TDGs) from Duc et al. (2000).
The straight line represents the empirical relationship found by Richer \& McCall (1995) for their 
dwarf sample. 

The majority of the Hercules cluster galaxies show metallicities following the empirical 
relationship for Irregular dwarfs, taking into account the 
uncertainties. Exceptions are ce-042, which looks overabundant for its luminosity and lies well 
within the loci of the LSBs and the TDGs, and ce-048, with a metallicity lower than expected 
for its luminosity and whose spectrum resembles that of typical BCDs.  
The presence of a few over--metallic object was also reported in the Hydra cluster by Duc et al. (2001a).
Unique among our objects in the Hercules cluster is the TDG ce-061 (Braine et al. 2001), 
whose dynamical nature as a gravitationally bound system inside the tidal tail of the IC 1182 
merger system has been confirmed recently through H$\alpha$ line Fabry-P\'{e}rot imaging 
(Duc \& Amram 2003). 

\subsubsection{The two tentative optical detections}  
Figures~\ref{tenta}(a) and~\ref{tenta}(b) show $V$-band contour plots of the two optical 
detections we tentatively associate with the VLA \HI\ sources sw-103 and sw-194. The 
filled circles indicate the centre positions of the \HI\ clouds. Although in both cases the difference 
in position between the centres of the optical and \HI\  sources are small compared to the VLA 
beam size (see below), which favours their association, the physical link between the \HI\ 
detections and the optical counterparts could not be proven unequivocally, as we could not obtain 
optical spectra for these two galaxies since they are too faint. Such optical spectra are needed to 
confirm their associations to the \HI\ clouds. 

Unfortunately, sw-103 is too close to a bright star for a detailed morphological analysis. The 
difference between the optical and \HI\ centre is only about 4$''$, or one fifth of the VLA HPBW. 
Although its $V$-band magnitude could be estimated ($m_{V} \approx 19.1$), this was not 
possible in the $I$-band due to the close bright star. 

The other tentative case, sw-194, may well be spurious. It is noted in D97 that this VLA \HI\ 
source could be an artefact due to a nearby strong continuum source (see Appendix A). It is 
clearly extended (about \am{1}{5}$\times$\am{0}{5}) and four galaxies (see Figure~\ref{tenta}(b)) 
lie within its contour. Our tentative identification concerns the brightest and largest of the four
objects only, which lies about 11$''$ from the centre of the VLA source. It is a very faint galaxy, 
for which we estimated $m_{V} \approx 20.6$ and $(V-I) \approx 1.8$, which is redder than expected 
for an \HI-rich galaxy. Although the three other objects are fainter and smaller, their association 
with the \HI\ source remains possible.

\section{Discussion}  
As mentioned earlier, one of our aims is the identification of \HI-selected dwarf galaxies 
in clusters. Although different dwarf selection criteria based on the \HI\ and optical properties 
of galaxies are used in the literature, there is no precise and universally accepted definition of 
what exactly is a dwarf galaxy (e.g., Binggeli 1994). 
The classification criteria we can apply to our data are:
narrow \HI\ line width, gaussian \HI\ line shape, faint optical luminosity and central surface brightness 
as function of luminosity. Two galaxies, ne-398 and 47-154, could not be classified, as they 
were not detected in \HI\ at Arecibo and therefore lack an accurate line width.

The only two systems in our Hercules cluster sample that satisfy all four dwarf selection criteria 
are ne-204 and ce-143: both have a $W_{50}$ profile width smaller than 100 \kms, a gaussian \HI\ 
proflle shape, an absolute blue luminosity fainter than -18 (see Table 4) and they lie among the 
dwarfs in Figure~\ref{dwarfsdiag}.

The levels of star formation in the four galaxies for which optical imaging and spectroscopy was 
obtained -- as indicated by their H$\beta$ equivalent widths and optical colours -- were found to 
be typical of those of active star forming galaxies. 

The metallicities of two objects do not follow the same trend as the five others: 
ce-042 was found to be overabundant for its luminosity and 
to lie well within the loci occupied by tidal dwarfs and LSBs, whereas ce-048 shows 
a very low metallicity for its luminosity. In fact, it is likely that the presence of a strong burst of star 
formation in ce-48, as its large H$\beta$ equivalent width suggests, has raised its luminosity, 
as proposed by Papaderos et al. (1996b), thus shifting it significantly in the $B$-band luminosity 
vs. metallicity diagram. 

The presence of two gas clouds with tentative optical detections (sw-103 and sw-194) should also be 
noted. While the former was confirmed 
in \HI\ at Arecibo, the latter, which could not be observed at Arecibo, could be a spurious VLA detection, 
according to D97. If the optical counterparts are really physically associated to the neutral gas clouds, 
they would be dwarf galaxies according to their optical magnitudes of 
$M_B$$\sim$ -16.9 and -14.8, respectively. If our tentative identification of a single galaxy with sw-194
were correct and the \HI\ source not spurious (see Section 3.2.4), then its \HI\ mass-to-blue luminosity 
ratio would be about 16 \MsunLBsun, an unprecedented high value for a dwarf, while sw-103 would 
have a quite mundane ratio of 1.1 \MsunLBsun. If the physical association between these \HI\ and 
optical sources were not confirmed, we may be dealing with gas clouds without detected optical 
counterparts in a cluster, whose existence would impose constraints on the cluster properties. 

Our accumulated data on the Hercules and Hydra cluster galaxies will be discussed further in a future 
paper on the influence of the cluster environment on galaxies, specifically dwarfs (Duc et al., 
in preparation).

\section{Conclusions}  
According to various diagnostics, two of our sample galaxies 
(ce-143 and ne-204) can be classified as dwarfs.
Of the galaxies studied in the optical, the star formation properties are similar to those of other 
samples of actively star forming galaxies and their metallicities are consistent with the 
blue luminosity-metallicity relation of nearby dwarfs, except for two galaxies, one of which is 
over-metallic and the other under-metallic for its luminosity. 
For two others the physical association with the \HI\ clouds seen 
superimposed on them could not be proven, since no optical redshifts could be obtained of them. If 
confirmed, one of them (sw-194) would have an extremely high \HI\ content.
A unique object in our sample is located in the IC 1182 merger system: the 
Tidal Dwarf Galaxy ce-061, which shows a rather large luminosity, metallicity and \HI\ content.

\begin{acknowledgements}
 Some of the data presented here have been taken using ALFOSC, which is owned by the Instituto 
de Astrof\'{\i}sica de Andaluc\'{\i}a (IAA) and operated at the Nordic Optical Telescope under 
agreement between IAA and the NBIfAFG of the Astronomical Observatory of Copenhagen. The 
NOT and WHT are operated on the island of La Palma by the NOT and ING groups, at the 
Observatorio del Roque de Los Muchachos of the Instituto de Astrof\'{\i}sica de Canarias. We also 
acknowledge the Service Program of the ING group for useful observations presented in this paper. 
We would like to thank the staff of Arecibo Observatory for their help with the observations and 
data reduction, especially P. Perrillat. IRAF, the Image Reduction and Analysis Facility, is written 
and supported at the National Optical Astronomy Observatory. We have made use of the 
NASA/IPAC Extragalactic Database (NED), which is operated by the Jet Propulsion Laboratory, 
California Institute of Technology, under contract with the National Aeronautics and Space 
Administration, and the Lyon-Meudon Extragalactic Database (LEDA). JIP acknowledges the Fifth 
Framework Program of the European Union for a Marie Curie Postdoctoral Fellowship. 
Research by PP has been supported by the Deutsches Zentrum f\"{u}r Luft-- und Raumfahrt e.V. 
(DLR) under grant 50\ OR\ 9907\ 7. C.B.F. 
\end{acknowledgements}



\newpage

\clearpage

\begin{table*}  
   \caption[]{Observations}
         \label{observations}
$$
\begin{tabular}{llrllll}
\hline
\vspace{-2 mm} \\
Name  & R.A.\,\,\,  &  Dec.\,\,\,  & $W_{50}$ & \HI\ det. & photom. & optical\\
           & VLA       & VLA & VLA & Arecibo &  & emission \\
 & \multicolumn{2}{c}{(J2000.0)} & (km/s) &  & & lines \\
\vspace{-2 mm} \\
\hline
\vspace{-1 mm} \\
ne-112  & 16 06 37.6 & 18 23 49 & 262 & yes &  &   \\
ne-142  & 16 06 22.5 & 18 00 03 & 171 & yes &  &   \\ 
ne-178  & 16 06 13.8 & 17 57 11 & 217 & yes &  &   \\ 
ne-204  & 16 06 05.9 & 18 09 20 & 125 & yes &  &   \\ 
ne-208  & 16 06 05.7 & 18 16 43 & 125 & yes &  &   \\ 
ne-240  & 16 05 58.1 & 18 24 41 & 262 & conf. &  &   \\ 
ne-250  & 16 05 52.2 & 18 27 58 & 307 & yes &  &   \\ 
ne-398  & 16 04 18.0 & 18 14 06 & 352 & weak &  &   \\  
ce-042  & 16 06 00.1 & 17 45 54 & 216: & yes & $B,V,i$ & yes \\
ce-048  & 16 05 55.7 & 17 42 39 & 171 & conf. & $B,V,i$ & yes \\  
ce-060  & 16 05 44.6 & 17 42 19 & 171 & conf. & $B,V,i$ & no \\  
ce-061 & 16 05 41.1 & 17 48 00 & 216 & conf & $B,V,i$ & yes \\  
ce-143 & 16 05 20.6 & 17 52 02 & 171 & yes & $V,i$ & yes \\ 
ce-176 & 16 05 09.9 & 17 51 20 &  171 & yes & $B,V,i$ & yes \\  
ce-200 & 16 05 06.7 & 17 47 00 &  216 & conf. & $B,V,i$ & yes \\  
sw-103 & 16 04 00.6 & 17 15 13 & 261 & yes & $B,V,i$ &    \\  
sw-194 & 16 01 07.1 & 17 20 16 & 395 & --- & $B,V,i$ &   \\  
sw-222 & 16 03 05.8 & 17 10 14 & 261 & yes & $V,i$ & yes \\
47-138  & 16 00 17.5 & 15 53 15 & 216 & --- &  &   \\ 
47-154  & 16 02 26.6 & 15 57 36 & 372 & no &  &   \\
47-166  & 16 02 16.2 & 16 04 41 &  261 & yes  &  &   \\ 
47-211  & 16 01 55.8 & 15 42 29 & 261 & yes  &  &   \\ 
\vspace{-2 mm} \\
\hline
\end{tabular}
$$
\begin{list}{}{}
\item[Arecibo \HI\ data: conf. indicates a spectrum confused by nearby galaxies,]
\item[--- objects for which no sensitive spectra could be obtained (see Section 3.X)]
\end{list}
\normalsize
\end{table*}

\newpage
\clearpage

\begin{table*}[!b]   
      \caption[]{Journal of the optical spectroscopic observations}
         \label{logspec}
$$
         \begin{tabular}{lccrr}
            \hline
            \noalign{\smallskip}
Object & Date & Telescope & P.A.$^{\mathrm{a}}$ & Width$^{\mathrm{b}}$ \\
            \noalign{\smallskip}
            \hline
            \noalign{\smallskip}
ce-042 & April 28 2001 & NOT & 5.6   & 1.0 \\
ce-048 & June  19 2000 & WHT & 336.4 & 1.0 \\
ce-060 & March 19 2001 & NOT & 319.3 & 1.0 \\
ce-061 & June  19 2000 & WHT & 96.6  & 1.1 \\
ce-143 & July  11 2000 & WHT & 51.5  & 1.0 \\
ce-176 & April 26 2000 & WHT & 317.5 & 1.1 \\
ce-200 & March 19 2001 & NOT & 70.7  & 1.0 \\
sw-222 & May   14 2001 & NOT & 81.0  & 1.0 \\
            \hline
            \noalign{\smallskip}
         \end{tabular}
$$
\begin{list}{}{}
\item[$^{\mathrm{a}}$] In degrees, from North towards East
\item[$^{\mathrm{b}}$] Width of the slit, in arcsecs.
\end{list}
\end{table*}

\newpage
\clearpage

\begin{table*}  
   \caption[]{Comparison of basic VLA and Arecibo \HI\  observational data}
         \label{hiressum}
$$
\begin{tabular}{llrrrrrrrrrr}
\hline
\vspace{-2 mm} \\
 & \multicolumn{4}{c}{----------- VLA data ------------} & 
 \multicolumn{5}{c}{------------------ Arecibo data --------------------} \\
Name  &  $V_{HI}$ & $W_{50}$ & $I_{H}$ & $I_{ext}$ &  $V_{HI}$ & 
  $W_{50}$ & $W_{20}$ & $I_{HI}$ & rms \\
 & km/s & km/s &   \multicolumn{2}{c}{Jy km/s} & km/s
  & km/s  & km/s & Jy km/s & mJy \\
\vspace{-2 mm} \\
\hline
\vspace{-1 mm} \\
     ne-112  & 11046 & 262 & 0.23 & 0.24 &  11014$\pm$6  & 156 & 188 & 0.49$\pm$0.09 & 0.63   \\  
     ne-142  & 11711 & 171 & 0.19 & 0.36 &  11726$\pm$5  & 128 & 153 & 0.35$\pm$0.07  & 0.47  \\  
     ne-178  & 11556 & 217 & 0.63 & 0.92 &  11591$\pm$4  & 191 & 242 & 1.10$\pm$0.09  & 0.49  \\  
     ne-204  & 11467 & 125 & 0.56 & 0.70 &  11451$\pm$3  &  62 &  93 & 0.54$\pm$0.06   & 0.60  \\  
     ne-208  & 11556 & 125 & 0.10 & 0.20 &   11521$\pm$7  & 218 & 244 & 0.25$\pm$0.07  & 0.39  \\  
     ne-240  & 11645 & 262 & 0.18 & 0.32 &  (11636$\pm$6  & 233 & 241 & 0.21) & 0.39  \\  
     ne-250  & 11645 & 307 & 0.88 & 1.32 &   11686$\pm$5  & 283 & 307 & 0.84$\pm$0.08  & 0.39  \\  
     ne-398  & 10602 & 352 & 0.50 & 1.45 &  10820$\pm$14 & 364 & 377 & 0.15$\pm$0.06  & 0.22  \\  
ce-042  & 11919: & 216: & 0.46: & 0.67: &  11910$\pm$3  & 127 & 135 & 0.33$\pm$0.07  & 0.54  \\  
ce-048  & 11145 & 171 & 0.20 & 0.36 &  (11134$\pm$2  & 208 & 211 & 0.53) & 0.61  \\  
ce-060  & 11100 & 171 & 0.27 & 0.41 & (11085$\pm$4  & 130 & 150 & 0.51) & 0.49  \\  
ce-061 & 10104 & 216 & 3.16 & 3.44 &  10263$\pm$3  & 595 & 666 & 3.44$\pm$0.18 & 0.76  \\  
ce-143 & 11587 & 171 & 0.16 & 0.31 &  11603$\pm$3  &  95 & 101 & 0.14$\pm$0.04  & 0.40  \\  
ce-176 &  9905 & 171 & 0.43 & 0.51 &  9919$\pm$4   & 150 & 183 & 0.65$\pm$0.08  & 0.56  \\  
ce-200 &  9927 & 216 & 0.10 & 0.17 & (9938$\pm$4:  & 177: & 189: & 0.46:) & 0.43  \\  
sw-103 & 11021 & 261 & 0.15 & 0.24 & 10999$\pm$4   & 155 & 159 & 0.18$\pm$0.06  & 0.38  \\  
sw-222 & 10093 & 261 & 0.25 & 0.49 & 10076$\pm$4   & 192 & 226 & 0.79$\pm$0.09  & 0.50  \\  
     47-154  & 10633 & 372 & 0.12 & 0.14 &    ---        &     &     &       & 0.32  \\  
     47-166  &  9849 & 261 & 0.45 & 0.62 &  9887$\pm$3   & 248 & 293 & 0.84$\pm$0.06  & 0.35  \\  
     47-211  & 10423 & 261 & 0.30 & 0.58 & 10460$\pm$3   & 280 & 295 & 0.83$\pm$0.10  & 0.51  \\  
\vspace{-2 mm} \\
\hline
\end{tabular}
$$
\begin{list}{}{}
\item[Arecibo \HI\ line parameters in brackets are for objects with profiles confused by nearby objects]
\end{list}
\normalsize
\end{table*}

\newpage

\clearpage

\begin{table*}[!h]  
      \caption[]{Basic properties of the sample galaxies}
         \label{propsum}
         \begin{tabular}{lcccccc}
            \hline
            \noalign{\smallskip}
Ident. & $V_{HI}$ & spec. & $M_{HI}$ & $M_{HI}/L_B$ & $W_{50}$ &
$M_{B}$ \\
 & km/s  & form$^{\mathrm{a}}$ & $10^{9}M_{\odot}$ &
$M_{\odot}$/$L_{\odot,B}$  & km/s & mag \\
            \noalign{\smallskip}
            \hline
            \noalign{\smallskip}
     ne-112  &   11014  &  LS  &  1.7  &  0.24  &  156   & $-19.1$  \\ 
     ne-142  &   11726  &  FT  &  1.8  &  3.7   &  128   & $-16.2$  \\
     ne-178  &   11591  &  DH  &  5.8  &  0.56  &  191   & $-19.5$  \\
     ne-204  &   11451  &  G   &  2.9  &  6.8   &   62   & $-16.0$  \\
     ne-208  &   11521  &  G   &  1.3  &  1.1   &  218   & $-17.1$  \\
     ne-240  &  (11645) &  --  & (1.7) & (1.5)  & (262)  & $-18.2$  \\
     ne-250  &   11686  &  FT  &  4.4  &  0.96  &  283   & $-19.8$  \\
     ne-398  &     ---  &  --- &  ---  &   ---  &  ---   & $-16.3$  \\
     ce-042 &   11910  &  DH  &  1.7  &  0.51  &  127    & $-17.2$  \\ 
     ce-048 &  (11145) &  --  & (1.9) & (3.5)  & (171)   & $-15.4$  \\ %
     ce-060 &  (11100) &  --  & (2.2) & (2.1)  & (171)   & ---      \\ 
     ce-061 &  (10104) &  --  & (18.2)& (7.3)  & (216)   & ---      \\ 
     ce-143 &   11603  &  G   & 0.74  &  0.35: &   95    & $-18.0$  \\ 
     ce-176 &    9919  &  LS  &  3.4  &  1.0   &  150    & $-18.3$  \\ 
     ce-200 &   (9938) &  --  & (0.90)& (0.43) & (216)   & $-17.6$  \\ 
     sw-103 &   10999  &  DH  &  0.95 &  1.1   &  155    & $-16.8$  \\
     sw-194 &  (10159) &  --  & (2.0) &  15.7: & (395)   & $-14.8$      \\ 
     sw-222 &   10076  &  G   & 4.2   &  0.35: &  192    & $-19.8$  \\ 
     47-138  &  (11284) &  --  & (1.2) &  0.30  & (216)  & $-18.4$  \\ 
     47-154  &  (10633) &  --- & (0.74)&   ---  & (372)  & ---      \\ 
     47-166  &    9887  &  DH  &   4.4 &  1.4   &  248   & $-15.0$  \\
     47-211  &   10460  &  DH  &   4.4 &  0.41  &  280   & $-19.8$  \\  
\vspace{-2 mm} \\
            \hline
             \noalign{\smallskip}
         \end{tabular}
\begin{list}{}{}
\item[$^{\mathrm{a}}$] The \HI\ spectrum forms are: DH  double horned, FT flat topped,]
\item[G gaussian, and LS lop-sided]
\end{list}
\end{table*}

\clearpage

\newpage

\begin{table*}  
      \caption[]{Absolute magnitudes and colours}
         \label{magcol}
$$
         \begin{tabular}{lccc}
            \hline
            \noalign{\smallskip}
Name & $M_{V}$ & $(B - V)$ & $(V - I)$ \\
            \noalign{\smallskip}
            \hline
            \noalign{\smallskip}
ce-042 & $-18.53$ & 0.17 & 0.29 \\
ce-048 & $-16.60$ & 0.28 & 0.55 \\
ce-060 & $-17.24$ & 0.23 & 1.06 \\
ce-061 & $-18.24$ & 0.23 & 0.00 \\
ce-143 & $-18.13\dagger$ & -- & 0.51 \\
ce-176 & $-18.72$ & 0.44 & 0.65 \\
ce-200 & $-18.12$ & 0.34 & 0.74 \\
sw-103 & $-16.74$ & --- & --- \\
sw-194 & $-15.24$ & --- & 1.80 \\
sw-222 & $-20.02\dagger$ & -- & 0.68 \\
\hline
            \noalign{\smallskip}
         \end{tabular}
$$
$\dagger$ An average $(B - V) = 0.3$ was assumed to estimate $M_{B}$.
\end{table*}

\clearpage

\newpage

\begin{table*}[t]  
      \caption[]{Extinction-corrected emission line flux ratios with respect to 
                 H$\beta$$\dagger$.}
         \label{propspec}
$$
         \begin{tabular}{lccccccc}
            \hline
            \noalign{\smallskip}
 & ce-042 & ce-048 & ce-061 & ce-176 & ce-200 & ce-143 & sw-222 \\
            \noalign{\smallskip}
            \hline
            \noalign{\smallskip}
\mbox{[O{\sc ii}]}~$\lambda$3727\AA & 6.149 & 0.932 & 2.475 & 3.062 & 6.234 & 3.505 & 3.044 \\
            &(1.108)&(0.050)&(0.122)&(0.343)&(0.956)&(0.322)&(0.154)\\
H$_{\beta}$~$\lambda$4861\AA & 1.000 & 1.000 & 1.000 & 1.000 & 1.000 & 1.000 & 1.000 \\
         &(0.071)&(0.012)&(0.016)&(0.052)&(0.042)&(0.037)&(0.016)\\
\mbox{[O{\sc iii}]}~$\lambda$4959\AA &  ---  & 1.424 & 0.940 & 1.004 &  ---  & 0.834 & 0.597 \\
            &  ---  &(0.040)&(0.033)&(0.086)&  ---  &(0.065)&(0.038)\\
\mbox{[O{\sc iii}]}~$\lambda$5007\AA & 2.955 & 3.559 & 2.753 & 2.287 & 1.878 & 2.730 & 1.867 \\
           &(0.299)&(0.078)&(0.069)&(0.154)&(0.135)&(0.142)&(0.054)\\
\mbox{[N{\sc ii}]}~$\lambda$6548\AA & 0.203 & ---  & 0.142 &  ---  & 0.086 &  ---  & 0.167 \\
           &(0.080)& --- &(0.009)&  ---  &(0.063)&  ---  &(0.015)\\
H$_{\alpha}$~$\lambda$6563\AA & 2.891 & 2.756 & 2.891 & 2.770 & 2.890 & 2.889 & 2.890 \\
          &(0.528)&(0.108)&(0.105)&(0.334)&(0.344)&(0.277)&(0.109)\\
\mbox{[N{\sc ii}]}~$\lambda$6584\AA & 0.466 & ---  & 0.332 & 0.298 & 0.186 & 0.447 & 0.407 \\
           &(0.116)& --- &(0.021)&(0.045)&(0.068)&(0.067)&(0.019)\\
\mbox{[S{\sc ii}]}~$\lambda$6717\AA &  ---  &  ---  & 0.402 & 0.529 &  ---  & 0.703 & 0.573 \\
           &  ---  &  ---  &(0.022)&(0.066)&  ---  &(0.096)&(0.026)\\
\mbox{[S{\sc ii}]}~$\lambda$6731\AA &  ---  &  ---  & 0.300 & 0.365 &  ---  & 0.513 & 0.371 \\
           &  ---  &  ---  &(0.017)&(0.047)&  ---  &(0.057)&(0.017)\\
\hline
$V_{hel}$$^{a}$ & 11840 & 11181 & 10012 & 9875 & 9940 & 11616 & 10087 \\
          &  (90) &  (39) &  (76) & (95) & (65) &  (63) &  (20) \\
$C$(H$_{\beta}$) & 0.298 & 0.000 & 0.004 & 0.000 & 0.103 & 0.116 & 0.166 \\
              &(0.117)&(0.025)&(0.023)&(0.077)&(0.076)&(0.061)&(0.024)\\
$E.W.$(H$_{\beta}$)$^{b}$ & 10.4 & 47.7 & 30.9 & 14.3 &  8.9 & 13.0 & 14.5 \\
                 &  (1.0) &  (2.5) &  (0.9) &  (1.1) &  (0.4) &  (0.6) &  (0.3) \\
            \hline
            \noalign{\smallskip}
         \end{tabular}
$$
\begin{list}{}{}
\item[$^{\mathrm{a}}$] In km~s$^{-1}$
\item[$^{\mathrm{b}}$] In \AA\
\item[$\dagger$] Quantities in parenthesis correspond to the
errors of the quantities quoted above
\end{list}
\end{table*}

\newpage

\clearpage

\begin{table*} 
\caption{Photometric results and structural properties of the galaxies
with definite optical counterparts$^a$}
\label{Tab1}
\begin{tabular}{lccccccccccc}
\hline
\hline
Name & Band & $\mu_{0}$ & $\alpha $ & $m_{\rm LSB}^{\rm fit}$ &
$P_{25}$  & $m_{P_{25}}$ & $E_{25}$ & $m_{\rm E_{25}}$  & $m_{\rm SBP}$ &
$r_{\rm eff}$,$r_{80}$ & $\eta$ \pano \\
     &     & \sbb\ &  kpc   & mag    & kpc       &  mag       &  kpc
          &  mag             &    mag & kpc & \\
    (1) &   (2)         &   (3)     &  (4)      &   (5)            &
 (6)    &  (7)             &  (8)     &  (9)  & (10) & (11) & (12) \kato \\
\hline
ce-042 & $B$ & 21.25$\pm$0.50  & 2.14$\pm$0.23 & 17.67 & 2.96  & 19.35  & 7.14  & 18.06  &  17.50  & 4.35,7.29 & 0.91 \pano  \\
{\sl (2.8,0.9)}  & $V$ & 21.26$\pm$0.45  & 2.24$\pm$0.22 & 17.59 & 3.48  & 18.91  & 7.87  & 17.98  & 17.31  & 4.35,7.36 & 0.91 \pano  \\
      & $I$ & 21.13$\pm$0.52  & 2.32$\pm$0.30 & 17.38 & 4.01  & 18.43  & 8.20  & 17.62  &  17.04  & 4.26,7.31 & 0.85 \pano  \\
\hline
ce-048  & $B$ & 22.68$\pm$0.16  & 1.17$\pm$0.06 & 19.61 & 1.06  & 22.25  & 2.49  & 20.11  & 19.52  & 1.65,3.06 & 0.89 \pano  \\
 $\star$ & $V$ & 22.59$\pm$0.15  & 1.19$\pm$0.06 & 19.48 & 1.55  & 20.79  & 2.64  & 19.94  & 19.21  & 1.48,2.87 & 0.77 \pano  \\
      & $I$ & 22.51$\pm$0.34  & 1.23$\pm$0.10 & 19.32 & 1.89  & 19.68  & 2.83  & 19.76  & 18.72  & 1.30,2.56 & 0.62 \pano  \\
\hline
ce-060 & $B$ & 23.42$\pm$0.13  & 2.45$\pm$0.13 & 18.74 & 0.89  & 23.00  & 3.56  & 19.67  & 18.85  & 3.26,5.38 & 0.90 \pano  \\
 $\star$  & $V$ & 23.15$\pm$0.10  & 2.41$\pm$0.10 & 18.51 & 1.41  & 19.14  & 4.10  & 19.25  & 18.59  & 3.07,5.21 & 0.87 \pano  \\
      & $I$ & 22.11$\pm$0.13  & 2.44$\pm$0.14 & 17.45 & 6.56  & 17.73  & 6.48  & 17.77  & 17.54  & 3.08,5.23 & 0.83 \pano  \\
\hline
ce-061 & $B$ & 22.68$\pm$0.27  & 2.41$\pm$0.19 & 18.72 & 3.87  & 18.46  & $<P_{25}$  & --  &  17.84  & 2.70,5.33 & 0.63 \pano  \\ 
{\sl (2.4,0.88)} & $V$ & 22.44$\pm$0.23  & 2.38$\pm$0.16 & 18.50 & 3.95  & 18.19  & 3.91  & 19.96  & 

 17.60  & 2.60,5.21 & 0.61 \pano  \\
      & $I$ & 21.50$\pm$0.42  & 2.19$\pm$0.25 & 17.72 & 3.91  & 17.80  & 6.93  & 18.12  &  17.06  & 2.69,5.52 & 0.63 \pano  \\
\hline
ce-143& $V$ & 21.32$\pm$0.11  & 1.20$\pm$0.04 & 18.2 & ---  & ---  & 4.08  & 18.36  & 18.32  & 2.07,3.48 & 1.09 \pano  \\
 $\star$   & $I$ & 20.97$\pm$0.13  & 1.24$\pm$0.05 & 17.78 & ---  & --- & 4.59  & 17.91  & 17.82  & 2.03,3.42 & 0.99 \pano  \\
\hline
ce-176  & $B$ & 21.89$\pm$0.20  & 2.26$\pm$0.10 & 17.69 & 2.01  & 20.39  & 6.45  & 18.06  &  17.54  & 4.22,6.87 & 1.17 \pano  \\
{\sl (2.1,0.7)}  & $V$ & 21.61$\pm$0.17  & 2.35$\pm$0.10 & 17.33 & 3.16  & 19.22  & 7.32  & 17.62  & 
 17.12  & 4.11,6.83 & 1.09 \pano  \\
      & $I$ & 21.11$\pm$0.31  & 2.41$\pm$0.18 & 16.77 & 4.90  & 17.99  & 8.63  & 16.96  &  16.46  & 3.96,6.71 & 1.00 \pano  \\ 
\hline
ce-200 & $B$ & 20.35$\pm$0.32  & 1.16$\pm$0.07 & 18.31 & 2.26  & 19.89  & 4.94  & 18.52  &  18.07  & 2.59,4.30 & 1.41 \pano  \\
{\sl (3.3,0.92)} & $V$ & 20.06$\pm$0.21  & 1.16$\pm$0.05 & 18.00 & 2.44  & 19.44  & 5.30  & 18.12  & 
 17.72  & 2.59,4.30 & 1.39 \pano  \\
      & $I$ & 19.35$\pm$0.09  & 1.18$\pm$0.03 & 17.27 & 2.67  & 18.67  & 6.24  & 17.36  &  17.00  & 2.59,4.37 & 1.23 \pano  \\ 
\hline
sw-222 & $V$ & 20.31$\pm$0.07  & 1.64$\pm$0.03 & 16.51 & ---  & ---  & 7.08  & 16.59  & 16.42  & 2.55,4.40 & 0.97 \pano  \\
$\star$   & $I$ & 19.65$\pm$0.08  & 1.65$\pm$0.04 & 15.83 & ---  & ---  & 8.13  & 15.88  & 15.74  & 2.53,4.42 & 0.91 \pano  \\
\hline
\hline
\end{tabular}
\parbox{17.2cm}{$a$: All listed values are corrected for extinction assuming a $C$(H$\beta$)
as listed in Table~\ref{propspec}}
\end{table*}

\begin{table*}   
      \caption[]{Oxygen abundances ($12 + \log$O/H)}
         \label{oxabun}
$$
         \begin{tabular}{lcccccc}
            \hline
            \noalign{\smallskip}
Name & \multicolumn{2}{c}{p-method} & \multicolumn{2}{c}{$R_{23}$} & N2Ha & Adopted \\
 & low & up & low & up & & abundance \\
            \noalign{\smallskip}
            \hline
            \noalign{\smallskip}
ce-042 & 8.52 & 7.91 & 8.49 & 8.24 & 8.68 & 8.68: \\
ce-048 & 7.58 & 8.61 & 7.73 & 8.71 & ---  & 7.58 \\
ce-061 & 7.88 & 8.41 & 7.98 & 8.62 & 8.53 & 8.41 \\
ce-176 & 8.02 & 8.35 & 8.03 & 8.60 & 8.48 & 8.35 \\
ce-200 & 8.66 & 7.94 & 8.43 & 8.31 & 8.27 & 8.27: \\
ce-143 & 8.10 & 8.26 & 8.14 & 8.51 & 8.66 & 8.26 \\
sw-222 & 8.03 & 8.37 & 7.99 & 8.64 & 8.62 & 8.37 \\
\hline
            \noalign{\smallskip}
         \end{tabular}
$$
\end{table*}


\newpage

\clearpage

\begin{figure*}  
 \centering
\caption{
Arecibo 21 cm \HI\ line spectra of 20 galaxies in the Hercules cluster;
for details on the ce-61 and ce-86 sources, see comments on ce-061 (Appendix~A).
The velocity resolution of the data is 9.1 \kms\ for most spectra, and 19.4 \kms\
for ne-398 and 47-154.
}
    \label{arecibospect}
 \end{figure*}



\begin{figure*}  
 \centering
\caption{
Comparison of $W_{50}$ line widths of integrated \HI\ profiles measured at Arecibo 
and the VLA, see the text for details.
}
    \label{w50comp}
 \end{figure*}



   \begin{figure*}  
   \centering
      \caption{
{\bf (a)} VLA \HI\ column density contours superposed on our optical $V$-band image
of field ce. The labels indicate the \HI-selected galaxies
in this field. R.A. and Dec. are in J2000. 
}
         \label{field}
   \end{figure*}

\addtocounter{figure}{-1}



   \begin{figure*}  
   \centering
      \caption{
{\bf (b)} VLA \HI\ column density contours superposed on our optical $V$-band image
of field sw. The labels indicate the \HI-selected galaxies
in this field. R.A. and Dec. are in J2000. 
}
   \end{figure*}



   \begin{figure*}
   \centering
      \caption{
Optical spectra of the seven galaxies showing emission lines.
}
         \label{spectra}
   \end{figure*}



\begin{figure*}[t]
\caption[]{Photometric properties of ce-042: {\bf Upper left:} $V$-band image with
superposed contours. The coordinates are in J2000.0 {\bf Lower left:} $(V - I)$
colour map with $V$-band contours superposed {\bf Upper right:} Surface brightness
profiles. The fitted surface brightness distribution of the LSB component in $B$ 
is depicted by the thick/grey curve. Open triangles show the emission in excess
of the LSB component {\bf Lower right:} Colour profiles
}
\label{ce042}
\end{figure*}



\begin{figure*}
\caption[]{Same as Figure~\ref{ce042} for ce-048}
\end{figure*}



\begin{figure*}
\caption[]{Same as Figure~\ref{ce042} for ce-060}
\end{figure*}



\begin{figure*}
\caption[]{Same as Figure~\ref{ce042} for
ce-061. The ellipse indicates the segment of tidal tail included in the surface
photometry analysis. The small crosses in the maps show a red pointlike sources
(probably a foreground star) which was removed before computing the SBPs.
}
\end{figure*}



\begin{figure*}
\caption[]{Same as Figure~\ref{ce042} for
ce-143. Due to strong overlapping with nearby objects, a 2-D model was fitted
and subtracted from the original images in order to be able to disentangle the
light of the \HI-selected galaxy. Large uncertainties are expected in the SBP
for $\mu_{V} > 25$~mag~arcsec$^{-2}$.}
\end{figure*}



\begin{figure*}
\caption[]{Same as Figure~\ref{ce042} for ce-176}
\end{figure*}



\begin{figure*}
\caption[]{Same as Figure~\ref{ce042} for ce-200}
\end{figure*}



\begin{figure*}
\caption[]{Same as Figure~\ref{ce042} for sw-222}
\label{sw222}
\end{figure*}



   \begin{figure*}
   \centering
      \caption{
Comparison of the absolute blue magnitude and blue central surface brightness of different 
types of galaxies and stellar subsystems, see text for details. To the original version
of Binggeli (1994) the loci of the LSB galaxies from van der Hulst (1998) have been added,
as well as the objects with available CCD surface photometry from our samples of  \HI-selected 
galaxies in the Hercules and Hydra clusters.
}
         \label{dwarfsdiag}
   \end{figure*}



   \begin{figure*}
   \centering
      \caption{
Absolute blue magnitude as function of H$\beta$ equivalent width for the Hercules
and Hydra cluster samples. For comparison we have added the
emission line galaxies from Salzer et al. (1989), represented by small
crosses. The straight line corresponds to the upper envelope from V\'{\i}lchez (1995).
}
         \label{salzer}
   \end{figure*}



   \begin{figure*}
   \centering
      \caption{
$(B - V)~vs.~(V - I)$ diagram for the Hercules cluster galaxies. Also shown are
ellipticals from Goudfrooij et al. (1994), spirals from Heraudeau \&
Simien (1996), amorphous galaxies from Gallagher \& Hunter (1987) and nearby dwarfs from
Makarova (1999).
}
         \label{colour}
   \end{figure*}

\newpage

\clearpage

   \begin{figure*}
   \centering
      \caption{
Metallicity~vs.~luminosity relationship for the Hercules and Hydra
cluster galaxies, represented by open diamonds and triangles respectively. 
The typical uncertainty in the metallicity determinations is
represented by the error bar at the bottom left.
Also shown are the sample of dwarfs from Richer \& McCall
(1995), the LSBs from van der Hulst et al. (1998) and the tidal dwarfs
from Duc et al. (2000), represented by crosses, squares and asterisks, respectively.
}
         \label{lumi_meta}
   \end{figure*}



\begin{figure*}
\caption{{\bf (a)} $V$-band contour plot of the tentative detection corresponding to sw-103.
The filled circle shows the centre position of the VLA H{\sc i} source from Dickey (1997).}
\label{tenta}
\end{figure*}



\addtocounter{figure}{-1}

\begin{figure*}
\caption{{\bf (b)} $V$-band contour plot of the tentative detection corresponding to sw-194;
we have assumed that the gas cloud is associated with the brightest galaxy in the
field, 11$''$ W of the centre position (see Sections 3.2.4 and 4).
The filled circle shows the position of the VLA H{\sc i} source from Dickey (1997).}
\end{figure*}

\newpage

\clearpage

\appendix

\section{Comments on individual objects}
Included are comments regarding the \HI\ properties,
morphological appearance and environment of the sample galaxies.

{\sl ne-112:}\, Although there is nearby galaxy detected in \HI\ at a similar redshift, ne-176 at 
\am{5}{8} separation, no significant confusion is expected of the Arecibo \HI\ spectrum. 
The D97 VLA data on ne-176 show $V_{HI}$= 11,101 \kms,  $W_{50}$= 508 \kms and $I_{ext}$= 0.81 Jy\kms. 
In our spectrum its flux density contribution is expected to be at most a negligible $\sim$12\% of the 
average flux density of ne-112.

{\sl ne-178:}\, The uncertain optical redshift of 11,688$\pm$251 \kms\ listed in LEDA is based on two
rather different measurements, 11,440 \kms\ (Hopp et al. 1995) and 11,935 \kms\ (Tarenghi et al. 1994).
The VLA and Arecibo \HI\ values of, respectively, 11,556 and 11,591 \kms\ are rather closer to 
the former optical value. The VLA data show a clear velocity gradient from the SW to the NE.

{\sl ne-204:}\, The VLA data show a highly extended \HI\ distribution, but no mention is made of
a velocity gradient in D97.

{\sl ne-208:}\, The VLA detection is unresolved, with only 13 pixels above threshold, i.e. an area of 
about half 
a beam size.

{\sl ne-240:}\, There are two nearby galaxies detected in \HI\ with redshifts similar to that of ne-240, 
ne-250, at \am{3}{5} separation, and ne-208, at \am{5}{8} separation, which are expected to cause strong 
confusion with the profile of ne-240. Except for the 3.6 times higher $I_{ext}$ line flux of ne-250, the 
VLA profile parameters of ne-240 and ne-250 are indistinguishable, seen the large velocity resolution, while 
ne-208 has a 90 \kms\ lower velocity and 0.6 times the VLA flux of ne-240. At most, we expect the average flux 
density of nearby ne-250 and ne-208 in our spectrum to be, respectively, $\sim$125\% and $\sim$30\% that 
of ne-240.

{\sl ne-250:}\, Its \HI\ spectrum is not expected to be confused significantly by that of nearby ne-240 
(see above), whose line emission is expected to be at most a negligible $\sim$11\% of the average flux 
density of ne-250.  The VLA data show an extended \HI\ distribution with a clear velocity gradient from 
the NW to the SE.

{\sl ne-398:}\,  This reported VLA source was not confirmed by the Arecibo data. The VLA detection 
has a peak line flux density of 2.1 mJy in the integrated profile corresponding to the $I_H$ integrated line 
flux (see Section 3.2), while the estimated mean flux density of the profile corresponding to the 
$I_{ext}$  line flux is 4.1 mJy. Such a line should have been easily detectable with the 0.22 mJy 
noise level of our Arecibo data. However, instead of a strong, 352 \kms\ wide profile centred on 
10,602 \kms, at Arecibo we only marginally detected a 364 \kms\ wide profile centred on  10,820 \kms, 
with a  3.2$\sigma$ peak after smoothing to a resolution of 19.5 \kms. Although the reported VLA source 
lies at the edge of the `ne' field, where the primary beam attenuation factor is 4, its detection 
seemed real, according to D97. Our Arecibo data indicate it was spurious.

{\sl ce-042:}\, The published VLA profile parameters are uncertain as the detection lies just
at the edge of the band. The VLA data show a velocity gradient from north to south.
Our optical redshift of 11840$\pm$90 \kms\ agrees within the uncertainties
with the 11959$\pm$60 \kms\ measured by Huchra et al. (1995).
The colour gradient was found to be 
$\gamma_+ \approx 0$~mag kpc$^{-1}$ in both $B-V$ and $V-I$ for $R^*\geq$2$''$.
Mean $B-V$ and $V-I$ colours of 0.18 and 0.27~mag,
respectively. The optical appearance of ce-042 is that of an Irregular
galaxy.
This object is almost isolated, with the elliptical galaxy
PGC~057123 ($B_{T}= 16.81$) as its closest bright neighbour, at a
distance of about 4$'$ and with a radial velocity difference of 1000~km~s$^{-1}$.
 
{\sl ce-048:}\, There are two nearby galaxies detected in \HI\ with redshifts similar to that of ce-048, 
which are expected to cause confusion with its Arecibo profile: ce-060, at \am{2}{7} separation, and ce-095, 
at \am{6}{8} separation. 
There is only $\pm$one VLA velocity channel (44 \kms) difference between their central \HI\ velocities 
and that of ce-048; the VLA line widths of ce-048 and ce-060 are the same, 171 \kms, while that of ne-95 is 
twice as large. We expect the average flux density of nearby ce-060 and ce-095 in our spectrum to be, at 
most, respectively, $\sim$66\% and $\sim$50\% of that of ce-048.
The VLA data show a  quite extended distribution, with a velocity gradient from NW to SE.
Its $\gamma_+\!\!\approx$=--0.05 \cg\ in both colours, and its
mean $B-V$ and $V-I$ colours are 0.27 and 0.38~mag,
respectively. It appears as a faint blue galaxy, more
compact than ce-042. There are several cluster galaxies near this
object, the closest one being PGC~057115 ($B_{T}= 16.84$), 
a lenticular galaxy with 400~km~s$^{-1}$ difference in radial
velocity, \am{1}{1} away.

{\sl ce-060}:\, Like in the case of ce-048 (see above), also for this pointing position
serious confusion is expected at Arecibo between the \HI\ lines of ce-048, ce-060 and ce-095. 
We expect the average flux density of nearby ce-048 and ce-095 in our spectrum to be, at 
most, respectively, $\sim$50\% and $\sim$70\% of that of ce-060.
Its $\gamma_+\!\!\approx$=--0.04 \cg\ in $B-V$ (mean colour 0.27~mag), and it shows practically 
no $V-I$ colour gradient, with a mean $V-I$ colour of 1.04~mag. 
In its vicinity lies the bright 
Sab galaxy IC~1185 ($B_{T}= 14.89$), at a distance of \am{0}{7}, with a velocity difference between 
both galaxies of about 700~km~s$^{-1}$.

{\sl ce-061}: An extended \HI\ distribution, measuring at least \am{2}{3} in the E-W direction, surrounds 
the peculiar galaxy IC 1182 ($B_{T}$=15.37, $V_{opt}$=10,223 \kms), formed to the East by the VLA source ce-061 and 
to the West by the much weaker source ce-086. In the optical, IC 1182 shows a jet-like structure towards the East, 
following the direction of the \HI\ distribution. The target galaxy ce-061 lies at the tip of the \HI\ tail 
extending eastwards from IC~1182, at about \am{1}{5} from the centre of IC 1182. 
The CCD image of ce-061 (Figure 8)shows two distinct peaks, and the maximum in the \HI\ tail coincides, within the 
25$''$ beam size, with the easternmost optical peak. The SBP has been taking over the area covering both
concentrations seen in Figure 8.
A recent H$\alpha$ line Fabry-P\'{e}rot velocity field (Duc \& Amram 2003) shows that
ce-061 is a gravitationally bound system inside the tidal tail of the IC 1182 merger system,
confirming its interpretation as a tidal dwarf galaxy by Braine et al. (2001), who detected about 
7 10$^9$M$_{\odot}$ of H$_2$ in a resolved distribution in IC 1182, but failed to detect ce-061 in the 
CO(1-0) or (2-1) lines, putting an upper limit to its H$_2$ mass of about 6 10$^7$M$_{\odot}$.
We obtained Arecibo \HI\ spectra of ce-061 and ce-086, with pointing centres corresponding to the VLA 
\HI\ positions, which are \am{1}{8} (half the Arecibo HPBW) apart. Though the two peaks
that can be identified with the VLA profiles of ce-061 and ce-086 are clearly present in our spectrum,
centered on $\sim$10,080 and 10,450 \kms, respectively, the latter component is much stronger than
in the VLA profiles: the average flux density in both components is about 16 and 2 mJy at the VLA, 
using the $I_{ext}$ fluxes, while their peak fluxes are $\sim$15 and 6 mJy in the Arecibo profile 
centered on ce-061 and  $\sim$9 and 6 mJy in the profile centered on ce-086. 
This may be due to extended emission in the distribution not included in the VLA profiles.
The VLA data show a fairly smooth velocity gradient along the major axis of the \HI\ distribution
and very little gas between the two peaks, at 10,200--13,000 \kms, in the area of IC 1182 itself.
$\gamma_+\!\!\approx$=--0.013 \cg\ in $B-V$ (mean colour 0.22~mag); $V-I$ colour gradient of 
0.016 \cg, mean colour 0.56~mag. 
\HI\ detections of IC 1882 were also reported by Bieging \& Biermann (1983), Bird et al. (1993), 
Bothun et al. (1981,1985), Mirabel \& Wilson (1984), Salpeter \& Dickey (1985) and 
Schommer et al. (1981). The average profile parameters from these references are 
$V_{HI}$= 10,239$\pm$62 \kms, $W_{20}$=  592$\pm$48  \kms\ and $I_{HI}$= 3.6$\pm$0.8 \Jykms. 
These values correspond well with our Arecibo parameters for ce-61 of 
10,263 \kms, 666 \kms\ and 3.4 \Jykms, respectively.

{\sl ce-143}: 
The CDD images show three, overlapping object. The peak of the VLA \HI\ 
distribution corresponds to the object in the middle, the weakest of the three. Before determining
its morphological component, two-dimensional model light distributions of its two neighbours
were subtracted from the images.
Zero colour gradient, mean $V-I$ colour of 0.48~mag. The
optical counterpart of this galaxy seems to be a triple nucleated source, with
the \HI\ cloud centred on the middle one. 
At a distance of \am{0}{7} of this system we find the Sa galaxy PGC~057055, with $B_{T}= 15.88$,
which shows a radial velocity more than 700~km~s$^{-1}$ higher than ce-143.

{\sl ce-176}: Although the VLA data show a galaxy in the vicinity with a similar redshift,
ce-200, we do not expect our Arecibo profile to be confused with it: at \am{4}{4} distance, 
it has only a 22 \kms\ higher central velocity than ce-176, but its mean VLA line flux is three 
times smaller. We expect the average flux density of ce-200 in our spectrum to be, at 
most, a negligible 7\% of that of ce-176. The VLA data show a very extended \HI\ distribution.
The galaxy shows practically a zero gradient in both colour profiles, and 
mean $B-V$ and $V-I$ indices of 0.42~mag and 0.66~mag, respectively. This galaxy 
appears like a dwarf Irregular galaxy with a very extended \HI\ disc. Also, some knots of star formation 
are seen in the optical images. 
The closest bright companion of this galaxy, at a distance of \am{2}{5}, is PGC~057020 ($B_{T}= 16.22$), 
an SB0/Sa galaxy with a radial velocity difference of about 1000~km~s$^{-1}$ with respect to ce-176.

{\sl ce-200}: The Arecibo spectrum of this object shows three peaks: one around 9780 \kms\, another 
around 9940 \kms\ and an about 300 \kms wide feature around $\sim$10,200 \kms.  
Though the centre peak is very close in velocity to ce-200, it is expected to be seriously confused 
with emission from nearby ce-176 (see above). We expect the average flux density of ce-176 in our 
spectrum to be, at most, equal to that of ce-200.
No object can be found in the VLA data that might cause the peak at 9780 \kms, while the third
peak appears due to ce-199, at \am{1}{6} distance, for which the VLA data show a $V_{HI}$ of
10,204 \kms, a FWHM of 284 \kms and a mean flux density of 1.5 mJy.
$\gamma_+\!\!\approx$0 \cg\ in $B-V$; mean colour 0.34~mag. The $V-I$ profile shows a slight 
colour gradient of $\approx$0.03 \cg\ and a mean
value of 0.77~mag. No morphological peculiarities are apparent for this
galaxy. Although it shows no signs of tidal interaction, two bright galaxies appear close to it, 
showing also a similar radial velocity: NGC~6043A (V = 9879 \kms\ and $B_{T}= 14.10$) 
and NGC~6045 (V = 9986 \kms\ and $B_{T}= 14.87$), classified, respectively, as S0 and 
Sc, and located at distances of \am{1}{3} and \am{1}{6} from ce-200.

{\sl sw-103}: the VLA data show a velocity gradient from northeast to southwest. 
See also Sections 3.2.4 and 4.

{\sl sw-194: }\, The strong (713 mJy), extended continuum source in the nearby galaxy 
NGC 6034 made sensitive Arecibo line observations impossible and the VLA detection tentative. 
The VLA profile is weak (0.75 mJy peak in the $I_{H}$ profile) and there are only 15 pixels above 
the detection threshold. The ``line'' could be a figment of imperfect bandpass calibration and 
continuum subtraction, as the continuum is sufficiently strong to increase the noise in the 
spectral baselines in its vicinity (D97). 
See also Sections 3.2.4 and 4.

{\sl sw-222}: The VLA data show an extended \HI\ distribution, with a plume extending about
1$'$ towards the SE from the centre, and a velocity gradient along the main body from west
to east. $\gamma_+\!\!\approx$0 \cg\ in $V-I$; mean colour 0.68~mag. 
This galaxy shows an elongated shape with an \HI\ plume extending towards the South.
The closest bright galaxy, which is located 4~arcmin away, is the lenticular
PGC~056824 ($B_{T}= 14.00$), whose velocity difference with sw-222 is about
300~km~s$^{-1}$.

{\sl 47-138 }\, No sensitive Arecibo observations could be obtained due to the presence of 
the 1.2 Jy radio source 4C+15.53, whose centre is just 40$''$ offset from the galaxy. 
It could be detected at the VLA as the continuum source has a diameter of about 100$''$ 
and the VLA peak flux density, 50 mJy/beam, is low 
enough not to disturb the \HI\ detection (1.1 mJy peak in the $I_{H}$ profile).
The VLA data show a velocity gradient from north to south.

{\sl 47-154 }\,  This weak VLA source was not detected at Arecibo.
The VLA detection has a peak line flux of 0.7 mJy in the integrated profile
corresponding to the $I_H$ integrated line flux (see Section 3.1), while the estimated mean flux density 
of the profile corresponding to the $I_{ext}$ line flux is 0.37 mJy. It is classified as a somewhat 
tentative detection in D97.
With an 0.32 mJy rms noise level in our Arecibo data, this source appears too weak for detection or 
confirmation.

{\sl 47-211 }\, The VLA data show a velocity gradient from NE to SW, along the major axis of
the galaxy.

\end{document}